\documentclass[preprint,showpacs,preprintnumbers,amsmath,amssymb]{revtex4}
\usepackage{epsfig}

\begin{document}

\title{Asymptotic expansion for the resistance between two
maximum separated nodes on a $M
\times N$ resistor network}

\author{N. Sh. Izmailian $^{1,2,3,4}$  and Ming - Chang Huang $^{1}$}

\affiliation{$^1$ Department of Physics, Chung-Yuan Christian
University, Chungli 320, Taiwan.}

\affiliation{$^2$ Institute of Physics, Academia Sinica, Nankang,
Taipei 11529, Taiwan.}

\affiliation{$^3$ Yerevan Physics Institute, Alikhanian Br. 2,
375036 Yerevan, Armenia}

\affiliation{$^4$ International Center for Advanced Study, Yerevan
State University, 1 Alex Manoogian St., Yerevan, 375025, Armenia}

\date{\today}

\begin{abstract}
We analyze the exact formulae for the resistance between two
arbitrary notes in a rectangular network of resistors under free,
periodic and cylindrical boundary conditions obtained by Wu [J.
Phys. A \textbf{37}, 6653 (2004)]. Based on such expression, we
then apply the algorithm of Ivashkevich, Izmailian and Hu [J.
Phys. A \textbf{35}, 5543 (2002)] to derive the exact asymptotic
expansions of the resistance between  two maximum separated nodes
on an $M \times N$ rectangular network of resistors with resistors
$r$ and $s$ in the two spatial directions. Our results is $
\frac{1}{s}R_{M\times N}(r,s)= c(\rho)\, \ln{S}+c_0(\rho,\xi)
+\sum_{p=1}^{\infty} \frac{c_{2p}(\rho,\xi)}{S^{p}} $ with $S=M
N$, $\rho=r/s$ and $\xi=M/N$. The all coefficients in this
expansion are expressed through analytical functions. We have
introduced the effective aspect ratio $\xi_{eff} =
\sqrt{\rho}\;\xi$ for free and periodic boundary conditions and
$\xi_{eff} = \sqrt{\rho}\;\xi/2$ for cylindrical boundary
condition and show that all finite size correction terms are
invariant under transformation $\xi_{eff} \to {1}/\xi_{eff}$.
\end{abstract}
 \maketitle \vskip 0.2 cm

\vskip 1cm
\section{Introduction}
\label{introduction}
The calculation of the resistance between arbitrary node of
infinite networks of resistors is a well studied subject
\cite{resistor1,resistor2,resistor3}. Resistor networks have been
widely studied as models for conductivity problems and classical
transport in disordered media
\cite{resistor4,resistor5,resistor6}.

Besides being a central problem in electric circuit theory, the
computation of resistances is also relevant to a wide range of
problems ranging from random walks (see \cite{resistor2} and
\cite{Lovasz1996}, and discussions below), first-passage processes
\cite{Redner2001}, to lattice Green's functions
\cite{Katsura1971}. Little attention has been paid to finite
network, even though the latter are those occurring in real life.
Recently, Wu \cite{Wu2004} has revisited the two-point resistance
problem and deduced a closed-form expression for the resistance
between arbitrary two nodes for finite networks with resistors $r$
and $s$ in the two spatial directions. Later, Jafarizadeh, et.al.
\cite{Jafar2007} proposed an algorithm for the calculation of the
resistance between two arbitrary nodes in an arbitrary
distance-regular networks. However, the exact expression obtained
in \cite{Wu2004} is in the form of a double summation whose
mathematical and physical contents are not immediately apparent.
Quite recently Essam and Wu based on the exact expression for the
resistance between arbitrary two nodes for finite rectangular
network obtained in \cite{Wu2004} has derived the asymptotic
expansion for the corner-to-corner resistance $(R_{M\times
N}(r,s))$ on an $M\times N$ rectangular resistor network under
free boundary conditions. For the case $M=N$ and $r=s=1$ they
computed the finite-size corrections to the corner-to-corner
resistance up to order $N^{-4}$:
\begin{equation}
R_{N\times N}(1,1) = \frac{4}{\pi} \log N + 0.077\,318 +\frac
{0.266\,070}{N^2}
 -\frac {0.534\,779}{N^4} + O\left(\frac 1 {N^6}\right) .\nonumber
\end{equation}
The computation of the asymptotic expansion of the
corner-to-corner resistance (in other word the resistance between
two maximum separated nodes) of a rectangular resistor network has
been of interest for some time, as its value provides a lower
bound to the resistance of compact percolation clusters in the
Domany-Kinzel model of a directed percolation \cite{Domany1984}.

In experiments and in numerical studies of model systems, it is
essential to take into account finite size effects in order to
extract correct infinite-volume predictions from the data.
 As soon as one has a
finite system one must consider the question of boundary
conditions on the outer surfaces or ``walls'' of the system. The
systems under various boundary conditions have the same per-site
free energy, internal energy, specific heat, etc, in the bulk
limit, whereas the finite size corrections are different. To
understand the effects of boundary conditions on finite-size
scaling and finite-size corrections, it is valuable to study model
systems. Therefore, in recent decades there have many
investigations on finite-size scaling, finite-size corrections,
and boundary effects for model systems
\cite{Blote,Cardy1,huetal,izmailian2002,izmailian2002a,izmailian2003,izmailian2007,EssamWu2009}.
Of particular importance in such studies are exact results where
the analysis can be carried out without numerical errors.

In this paper we will derive the exact asymptotic expansions for
 resistance between two maximum separated nodes on the
rectangular network under free, periodic and cylindrical boundary
conditions. We will show that the exact asymptotic expansion of
the resistance between  nodes of the network for all boundary
conditions can be written as
\begin{eqnarray}
\frac{1}{s}R_{M\times N}(r,s)&=& c(\rho)\, \ln{S}+c_0(\rho,\xi)
+\sum_{p=1}^{\infty} \frac{c_{2p}(\rho,\xi)}{S^{p}}
\label{RmnAsymptotic}
\end{eqnarray}
where $\rho = r/s$, $S = M N$ is the area of the lattice and $\xi
= M/N$ is the aspect ratio. The all coefficients in this expansion
are expressed through analytical functions. We will show that all
finite size correction terms are invariant under transformation
$\xi \to {1}/({\rho \; \xi})$ for free and periodic boundary
conditions and under transformation $\xi \to 4/({\rho \; \xi})$
for cylindrical boundary condition, which actually means that
$\xi_{eff}$
\begin{eqnarray}
\xi_{eff}&=&\xi \; \sqrt{\rho}
\qquad \mbox{for free and periodic b.c.} \label{effective}\\
\xi_{eff}&=&\xi \; \sqrt{\rho}/2 \qquad \mbox{for cylindrical
b.c.} \label{effectivecyl}
\end{eqnarray}
can be regarded as the effective aspect ratio.

The organization of this paper is as follows: Based on the exact
expression for the resistance between arbitrary two nodes for
finite rectangular network under free, periodical and cylindrical
boundary conditions obtained in \cite{Wu2004} we express the
resistance between two most separated nodes in terms of
$G_{\alpha, \beta}(\rho,\xi)$ with $(\alpha, \beta)= (1/2, 0)$ and
$(0, 1/2)$ (Sec. II). We then extend Ivashkevich, Izmailian and Hu
algorithm \cite{izmailian2002}  to derive the exact asymptotic
expansions of the resistance between two maximum separated nodes
on the rectangular network  for all boundary conditions and write
down the expansion coefficients up to the second order (Sec. III).
We also discuss our results in Sec. IV.

\section{Two-dimensional resistor networks}
\label{Second part}

An electrical network can be regarded as a graph in which the
resistance $R_{ij}$ is associated to the edge between pair of
connected nodes i and j. Denote the electric potential at the i-th
vertex by $V_i$ and the net current flowing into the network at
the i-th vertex by $I_i$. When the potential difference occurs
between points i and j, the current is given by the Ohm's law
$I_{ij} = (V_i-V_j)C_{ij}$, where $C_{ij} = 1/R_{ij}$ is the
conductance of the respective link. By the Kirchhoff's current law
total current outflow from any point in the interior is zero,
$\sum_j I_{ij} = 0$, we then find for the voltage
\begin{equation}
V_i =\sum_j V_j C_{ij}/C_i \label{Kirch}
\end{equation}
where $C_i = \sum_j C_{ij}$ and the sum is over all nodes j which
are connected to i.

The two-point resistance has a probabilistic interpretation based
on classical random walker walking on the network. The averaging
property expressed by equation (\ref{Kirch}) implies that the
voltage is a harmonic function on the interior points of the
graph. This makes the basis for the probabilistic interpretation
of the voltage \cite{Ballobas1998,resistor2,Kemeny1976,Kelly1979}.
The random walk determined by the electrical network is defined as
finite state Markov chain (for more details see
\cite{resistor2})with the transition probabilities $P_{ij}$ that
are weighted with the conductances as $P_{ij} = C_{ij}/C_i$. Then,
when the constant voltage is applied to the graph such that $V_a =
1$ and $V_b = 0$, the voltage in an interior point x is determined
as the hitting probability $h_x$ that a walker staring at x
reaches the point a before reaching b.

Consider a rectangular $M \times N$ network of resistors with
resistances r and s on edges of the network in the respective
horizontal and vertical directions. The closed-form expression for
the resistance $R_{\{M\times N\}}({\bf r}_1,{\bf r}_2)$ between
arbitrary two nodes ${\bf r}_1=(x_1, y_1)$ and ${\bf r}_2=(x_2,
y_2)$ for free, periodic and cylindrical boundary conditions was
obtained in \cite{Wu2004}.

In what follows, we will show that the resistance $R_{\{M\times
N\}}(r,s)$ between two maximum separated nodes of the network for
all above mentioned boundary conditions can be expressed in terms
of $G_{\alpha, \beta}(\rho,M,N)$ only,
\begin{eqnarray}
R_{M\times N}^{\,\rm free}(r,s) &=&-r+\frac{\sqrt{r
s}}{S}\left(G_{0,1/2}(\rho,M,N)+G_{1/2,0}(\rho,M,N)\right)
,\label{2dRfin}\\
R_{M\times N}^{\rm per} (r,s) &=& \frac{\sqrt{r
s}}{S}\left(G_{0,1/2}(\rho,M/2,N/2)+G_{1/2,0}(\rho,M/2,N/2\right),
 \label{RRfin}\\
R_{M\times N}^{\rm cyl}(r,s) &=&\frac{\sqrt{r
s}}{S}\left(G_{0,1/2}(\rho,M/2,N)+G_{1/2,0}(\rho,M/2,N)\right),
\label{cylRfin}
\end{eqnarray}
where $G_{\alpha, \beta}(\rho,\xi)$ is given by
\begin{equation}
G_{\alpha, \beta}(\rho,M,N) =
M\;{\tt~Re}\sum_{n=0}^{N-1}f\left(\pi\frac{n+\alpha}{N}\right)\;
{\rm coth}\left[M\,\omega\left(\pi\frac{n+\alpha}{N}\right)+i \pi
\beta\right]\label{Gab}
\end{equation}
for $(\alpha, \beta) \ne (0,0)$. The function $\omega(x)$ is the
same for all boundary conditions and given by:
\begin{equation}
\omega(x) = {\rm arcsinh}\sqrt{\rho} \sin x \label{omega}
\end{equation}
and function $f(x)$ is depend on boundary conditions and given by
\begin{eqnarray}
f(x)&=&\frac{\cos^2 x\sqrt{1+\rho\sin^2x}}{\sin x}
\qquad \quad \mbox{for free BCs,} \label{fxfree}\\
f(x)&=&\frac{1}{\sin x \,\sqrt{1+\rho\sin^2x}} \qquad \quad \mbox{for periodic BCs,} \label{fxtor}\\
f(x)&=&\frac{\cos^2 x}{\sin x \,\sqrt{1+\rho\sin^2x}} \qquad \quad
\mbox{for cylindrical BCs} \label{fxcyl}
\end{eqnarray}

\subsection{Two-dimensional network: free boundary conditions}

\begin{figure}
\epsfxsize=90mm \vbox to2in{\rule{0pt}{2in}}
\includegraphics{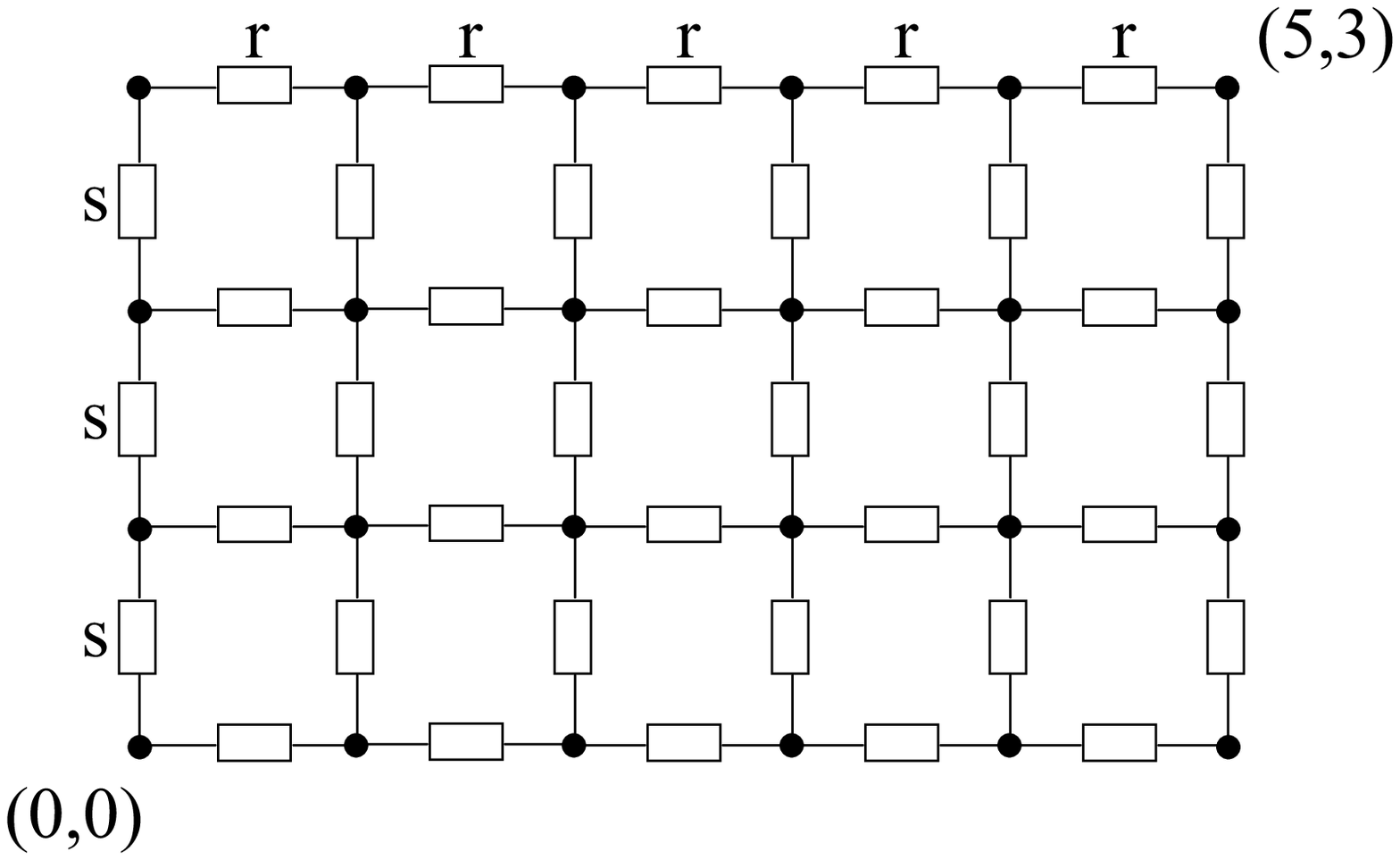}  \caption{A $6 \times 4$ rectangular network
with free boundary conditions}
\end{figure}

Consider a rectangular $M \times N$ network of resistors with free
boundary conditions and  with resistances r and s on edges of the
network in the respective horizontal and vertical directions. The
example of a rectangular network with $M = 6$, $N = 4 $ is shown
in Fig. 1. The resistance between the two maximum separated nodes
on the network of resistors with free boundary conditions is the
resistance between opposite corner nodes $(0, 0)$ and $(M-1,N-1)$
of the network which is given by \cite{EssamWu2009}
\begin{eqnarray}
R_{M \times N}^{free}(r,s)&=& \frac{r(M-1)}{N}+\frac{s(N-1)}{M}
\label{ResistanceFree0}\\
&+&\frac{2}{M
N}\sum_{m=1}^{M-1}\sum_{n=1}^{N-1}\frac{\left[\cos(\theta_m/2)\cos(\phi_n/2)
-\cos\left(M-\frac{1}{2}\right)\theta_m\cos\left(N
-\frac{1}{2}\right)\phi_n\right]^2}
{r^{-1}(1-\cos\theta_m)+s^{-1}(1-\cos \phi_n)} \nonumber
\end{eqnarray}
where $\theta_m=\pi m/M, \phi_n=\pi n/N$. With the help of the
identity
\begin{eqnarray}
\sum_{n=0}^{N/2-1}\cot^2 \frac{\pi (n+1/2)}{N} = \frac{N(N-1)}{2}
\label{ident40}
\end{eqnarray}
the Eq. (\ref{ResistanceFree0}) can be rewritten in the following
form
\begin{eqnarray}
R_{M \times N}^{free}(r,s)&=& -\frac{r(M-1)}{N}-\frac{s(N-1)}{M}
\label{ResistanceFree}\\
&+&\frac{2}{M
N}\sum_{m=0}^{M-1}\sum_{\substack{n=0\\
(m,n) \ne (0,0)}}^{N-1}\frac{\left[\cos(\theta_m/2)\cos(\phi_n/2)
-\cos\left(M-\frac{1}{2}\right)\theta_m\cos\left(N
-\frac{1}{2}\right)\phi_n\right]^2}
{r^{-1}(1-\cos\theta_m)+s^{-1}(1-\cos \phi_n)} \nonumber
\end{eqnarray}
Using the fact that $\cos\left(M-\frac{1}{2}\right)\theta_m =
(-1)^m \cos(\theta_m/2)$ and $\cos\left(N-\frac{1}{2}\right)\phi_n
= (-1)^n \cos(\phi_n/2)$ the Eq. (\ref{ResistanceFree}) can be
rewritten in the following form
\begin{eqnarray}
R_{M \times N}^{free}(r,s)&=& -\frac{r(M-1)}{N}-\frac{s(N-1)}{M}
\label{ResistanceFree1}\\
&+&\frac{4 r}{M N} \sum_{m=0}^{M-1}\sum_{\substack{n=0\\
m+n=odd}}^{N-1} \frac{\cos^2(\phi_n/2)\left(1+\rho
\sin^2(\phi_n/2)\right)}
{\sin^2(\theta_m/2)+\rho \sin^2(\phi_n/2)} \nonumber\\
&-&\frac{4 r}{M N} \sum_{m=0}^{M-1}\sum_{\substack{n=0\\
m+n=odd}}^{N-1} \cos^2(\phi_n/2)  \nonumber
\end{eqnarray}
There are two possibilities for the restriction $m+n =$ odd to
hold, namely, m-odd, n-even and m-even, n-odd. Splitting the sum
into two parts accordingly we obtain
\begin{eqnarray}
R_{M \times N}^{free}(r,s)&=& -\frac{r(M-1)}{N}-\frac{s(N-1)}{M}
\label{ResistanceFree11}\\
&+&\frac{4 r}{M N}\sum_{m=0}^{M/2-1}\sum_{n=0}^{N/2-1}\left[f(m+1/2,n)+f(m,n+1/2)\right]
\nonumber\\
&-&\frac{4 r}{M N} \sum_{m=0}^{M/2-1}\sum_{n=0}^{N/2-1}
\cos^2\frac{\pi n}{N}-\frac{4 r}{M N}
\sum_{m=0}^{M/2-1}\sum_{n=0}^{N/2-1} \cos^2\frac{\pi (n+1/2)}{N}
\nonumber
\end{eqnarray}
where
\begin{equation}
f(m,n)=\frac{\cos^2\frac{\pi n}{N}\left(1+\rho \sin^2\frac{\pi
n}{N}\right)} {\sin^2\frac{\pi m}{M}+\rho \sin^2\frac{\pi n}{N}}.
\label{transf1}
\end{equation}
Sums of the term $\cos^2(x)$ can be carried out using the
identities
\begin{eqnarray}
\sum_{n=0}^{N/2-1}\cos^2 \frac{\pi n}{N} =
\frac{N}{4}+\frac{1}{2}, \qquad \sum_{n=0}^{N/2-1}\cos^2 \frac{\pi
(n+1/2)}{N} = \frac{N}{4} \label{ident2}
\end{eqnarray}
This yields
\begin{eqnarray}
R_{M \times N}^{free}(r,s)&=& -\frac{r(M+N)}{N}-\frac{s(N-1)}{M}
\label{ResistanceFree12}\\
&+&\frac{4 r}{M
N}\sum_{m=0}^{M/2-1}\sum_{n=0}^{N/2-1}\left[f(m+1/2,n)+f(m,n+1/2)\right]
\nonumber
\end{eqnarray}
Now we first express double sums $\sum_{n=0}^{N/2-1}
\sum_{m=0}^{M/2-1}f(m,n)$ in terms of $\sum_{n=0}^{N-1}
\sum_{m=0}^{M-1}f(m,n)$, It it easy to show that
$f(m,N-n)=f(M-m,n)=f(m,n)$ and thus
\begin{eqnarray}
&2&\sum_{n=0}^{N/2-1}\sum_{m=0}^{M/2-1}\left[f(m+1/2,n)+f(m,n+1/2)\right]
=\frac{1}{2}\sum_{n=0}^{N-1}\sum_{m=0}^{M-1}\left[f(m+1/2,n)+f(m,n+1/2)\right]
\nonumber\\
&-&\sum_{n=0}^{N/2-1}\left[f(M/2,n+1/2)-f(0,n+1/2)\right]
-\sum_{m=0}^{M/2-1}\left[f(m+1/2,N/2)-f(m+1/2,0)\right]\label{transf2}.
\end{eqnarray}
With the help of the identities given by Eq. (\ref{ident40}) and
\begin{eqnarray}
\sum_{m=0}^{M/2-1}\frac{1}{\sin^2 \frac{\pi (m+1/2)}{M}}
=\frac{M^2}{2} \label{ident4}
\end{eqnarray}
The sums $\sum_{n=0}^{N/2-1}\left[f(M/2,n+1/2)-f(0,n+1/2)\right]$
, $\sum_{m=0}^{M/2-1}\left[f(m+1/2,N/2)+f(m+1/2,0)\right]$ can be
written as
\begin{eqnarray}
\sum_{n=0}^{N/2-1}\left[f(M/2,n+1/2)-f(0,n+1/2)\right]&=&
-\frac{1}{\rho}\sum_{n=0}^{N/2-1}\cot^2 \frac{\pi
(n+1/2)}{N}=-\frac{N(N-1)}{2\rho}
\nonumber\\
\sum_{m=0}^{M/2-1}\left[f(m+1/2,N/2)-f(m+1/2,0)\right]&=&-\sum_{m=0}^{M/2-1}\frac{1}{\sin^2
\frac{\pi (m+1/2)}{M}} = -\frac{M^2}{2}. \label{transf21}
\end{eqnarray}
Plugging Eqs. (\ref{transf2}) and (\ref{transf21}) back in Eq.
(\ref{ResistanceFree12}) we finally obtain
\begin{eqnarray}
R_{M \times N}^{free}(r,s)&=& -r+\frac{r}{M
N}\sum_{m=0}^{M-1}\sum_{n=0}^{N-1}\left[f(m+1/2,n)+f(m,n+1/2)\right]
\label{ResistanceFree1234}
\\
&=&-r+\frac{r}{M N}\sum_{n=0}^{N-1}\left(1+ \rho \sin^2\frac{\pi
n}{N}\right)\cos^2 \frac{\pi n}{N}\,\sum_{m=0}^{M-1} \Big[{\rho
\sin^2 \frac{\pi n}{N} +\sin^2 \frac{\pi
(m+\frac{1}{2})}{M}}\Big]^{-1}\nonumber \\
&+&\frac{r}{M N}\sum_{n=0}^{ N-1}\left(1+ \rho \sin^2\frac{\pi
(n+1/2)}{N}\right)\cos^2 \frac{\pi (n+1/2)}{N} \nonumber\\
&\times&\sum_{m=0}^{M-1} \Big[{\rho \sin^2 \frac{\pi (n+1/2)}{N}
+\sin^2 \frac{\pi m }{M}}\Big]^{-1}\nonumber
\end{eqnarray}
The sum over m in the Eq. (\ref{ResistanceFree1234}) can be
carried out using the identity \cite{GradshteinRyzhik}
\begin{equation}
\prod_{m=0}^{M-1}4\textstyle{ \left[~\!{\rm sinh}^2\omega +
\sin^2\frac{\pi (m+\beta)}{M}\right]} =4\left|~\!{\rm
sinh}\left(M\omega+i\pi\beta\right)\right|^2\label{identity11}
\end{equation}
Note that using more complicated approach the identity given by
Eq. (\ref{identity11}) has been obtained previously in
\cite{EssamWu2009} and \cite{Janke1}.

Taking the derivative over $\omega$ from the logarithm of the left
and right side of the equation (\ref{identity11}) we obtain
\begin{equation}
\sum_{m=0}^{M-1} \Big[{{\rm sinh}^2\omega+\sin^2
\frac{(m+\beta)\pi}{M}}\Big]^{-1}=2 M\,{\tt~Re} \frac{{\rm
coth}\left[M\,\omega +i \pi \beta\right]}{{\rm \sinh 2\omega}}.
\label{a3}
\end{equation}
Thus the sum over m in the Eq. (\ref{ResistanceFree1234}) can be
carried out as
\begin{equation}
\sum_{m=0}^{M-1} \Big[{\rho \sin^2 \frac{\pi n}{N} +\sin^2
\frac{\pi (m+\frac{1}{2})}{M}}\Big]^{-1}=2 M\,{\tt~Re} \frac{{\rm
coth}\left[M\,\omega(\frac{\pi n}{N})+i \pi/2\right]}{{\rm \sinh
2\omega(\frac{\pi n}{N})}} \label{a5}
\end{equation}
\begin{equation}
\sum_{m=0}^{M-1} \Big[{\rho \sin^2 \frac{\pi (n+1/2)}{N} +\sin^2
\frac{\pi m}{M}}\Big]^{-1}=2 M\,{\tt~Re} \frac{{\rm
coth}M\,\omega(\frac{\pi (n+1/2)}{N})}{{\rm \sinh
2\omega(\frac{\pi (n+1/2)}{N})}} \label{a51}
\end{equation}
where $\omega(x)$ is given by Eq. (\ref{omegaTaylor}). It is easy
to see that
\begin{equation}
{\rm \sinh 2\omega}(x) = 2 \sqrt{\rho}\sin{x}\sqrt{1+\rho
\sin{x}^2} \label{2omega}
\end{equation}

Plugging Eqs.  (\ref{a5}) and (\ref{a51}) back in Eq.
(\ref{ResistanceFree1234}) we obtain that $R_{M \times
N}^{free}(r,s)$ can be written in the form given by Eq.
(\ref{2dRfin}).

\subsection{Two-dimensional network: periodical boundary
conditions}
Consider a rectangular $M \times N$ resistor network
with periodic boundary conditions.  Using a closed-form expression
for the resistance between arbitrary two nodes for finite network
given by Eq. (43) of \cite{Wu2004} we can obtain for the
resistance between nodes $r_1=(0,0)$ and $r_2=(M/2,N/2)$ of the
network the following expression
\begin{eqnarray}R_{M\times N}^{\rm per}(r,s)
 =\frac{1}{M N}\,{\sum_{m=0}^{M-1}\sum_{\substack{n=0\\
(m,n) \ne (0,0)}}^{N-1} \frac {1
-\cos\left(M\theta_m+N\phi_n\right)} {r^{-1} (1-\cos 2\theta_m )
+s^{-1} (1-\cos 2\phi_n )} },\label{torR}
\end{eqnarray}
Using the fact that $\cos\left(M\theta_m+N\phi_n\right) =
(-1)^{m+n}$  the Eq. (\ref{torR}) can be rewritten in the
following form
\begin{eqnarray}
R_{M \times N}^{\rm per}(r,s)= \frac{r}{M N} \sum_{m=0}^{M-1}\sum_{\substack{n=0\\
m+n=odd}}^{N-1} \frac{1} {\sin^2\theta_m+\rho \sin^2\phi_n}
\label{tor1}
\end{eqnarray}
Splitting the sum into two parts accordingly we obtain
\begin{eqnarray}
R_{M\times N}^{per}(r,s)& =&\frac{r}{M N}
\sum_{m=0}^{M/2-1}\sum_{n=0}^{N/2-1} \frac{1}
{\sin^2\left(\frac{2\pi(m+1/2)}{M}\right)+\rho
\sin^2\left(\frac{2\pi n}{N}\right)}\nonumber\\
&+& \frac{r}{M N} \sum_{m=0}^{M/2-1}\sum_{n=0}^{N/2-1} \frac{1}
{\sin^2\left(\frac{2\pi m}{M}\right)+\rho \sin^2\left(\frac{2\pi
(n+1/2)}{N}\right)}\label{a2tor}
\end{eqnarray}
The sum over m in the Eq. (\ref{a2tor}) can be carried out using
the identities given by Eq. (\ref{a5}) and (\ref{a51}). This
yields
\begin{eqnarray}
R_{M\times N}^{per}(r,s)& =&\frac{\sqrt{r s}}{N} \sum_{n=0}^{
N/2-1}\frac{{\rm coth}\left[M\,\omega\left(\frac{2\pi
n}{N}\right)+i \pi/2\right]}{\sin \frac{2 \pi
n}{N}\sqrt{1+\rho\sin^2{\frac{2 \pi n}{N}}}}\label{a2tor1}\\
&+&\frac{\sqrt{r s}}{N} \sum_{n=0}^{ N/2-1}\frac{{\rm
coth}M\,\omega\left(\frac{2\pi (n+1/2)}{N}\right)}{\sin \frac{2
\pi (n+1/2)}{N}\sqrt{1+\rho\sin^2{\frac{2 \pi
(n+1/2)}{N}}}}.\nonumber
\end{eqnarray}
Introducing function $f(x)$ given by Eq. (\ref{fxtor}) we finally
arrived to the Eq. (\ref{RRfin}).

\subsection{Two-dimensional network: cylindrical boundary
conditions}
Consider a rectangular $M \times N$ resistor network
embedded on a cylinder with periodic boundary in the direction of
M and free boundaries in the direction of N. Using Eq. (46) of
\cite{Wu2004}, the resistance between nodes $r_1=(0,0)$ and
$r_2=(M/2,N-1)$ of the network is
\begin{eqnarray} &&R_{M\times N}^{\rm cyl}(r,s)
 =-\frac{r M}{4 N} \label{cylR} \\
&&+\frac{1}{M N}\,{\sum_{m=0}^{M-1}\sum_{\substack{n=0\\
(m,n) \ne (0,0)}}^{N-1} \Bigg(\frac
{\cos^2\left(\frac{1}{2}\phi_n\right)+\cos^2\left(N-\frac{1}{2}\right)\phi_n
 - 2\cos\left(\frac{1}{2}\phi_n\right)\cos\left(N-\frac{1}{2}\right)\phi_n \cos M\theta_m }
{r^{-1} (1-\cos 2\theta_m )  +s^{-1} (1-\cos \phi_n )} }\Bigg).
\nonumber
\end{eqnarray}
Using the fact that $\cos\left(M\theta_m\right) = (-1)^m$ and
$\cos\left(N-\frac{1}{2}\right)\phi_n = (-1)^n \cos(\phi_n/2)$ the
Eq. (\ref{cylR}) can be rewritten in the following form
\begin{eqnarray}
R_{M \times N}^{\rm cyl}(r,s)=-\frac{r M}{4 N}
+\frac{2 r}{M N} \sum_{m=0}^{M-1}\sum_{\substack{n=0\\
m+n=odd}}^{N-1} \frac{\cos^2(\phi_n/2)} {\sin^2\theta_m+\rho
\sin^2 (\phi_n/2)} \label{cyl1}
\end{eqnarray}
Splitting the sum into two parts accordingly we obtain
\begin{eqnarray}
R_{M \times N}^{\rm cyl}(r,s)&=&-\frac{r M}{4 N}+\frac{2 r}{M N}
\sum_{m=0}^{M/2-1} \sum_{n=0}^{N/2-1} \frac{\cos^2\frac{\pi n}{N}}
{\sin^2\frac{2\pi(m+1/2)}{M}+\rho \sin^2 \frac{\pi n}{N}}
\label{cyl2}\\
&+&\frac{2 r}{M N} \sum_{m=0}^{M/2-1} \sum_{n=0}^{N/2-1}
\frac{\cos^2\frac{\pi (n+1/2)}{N}} {\sin^2\frac{2\pi m}{M}+\rho
\sin^2 \frac{\pi (n+1/2)}{N}}\nonumber
\end{eqnarray}
Following along the same lines as in the case of free boundary
conditions the Eq. (\ref{cyl2}) can be written as
\begin{eqnarray}
R_{M \times N}^{\rm cyl}(r,s)&=&\frac{r}{M N}
\sum_{n=0}^{N-1}\cos^2\frac{\pi n}{N}\sum_{m=0}^{M/2-1} \frac{1}
{\sin^2\frac{2\pi(m+1/2)}{M}+\rho \sin^2 \frac{\pi n}{N}}
\label{cyl3}\\
&+&\frac{r}{M N} \sum_{n=0}^{N-1}\cos^2\frac{\pi (n+1/2)}{N}
\sum_{m=0}^{M/2-1}\frac{1} {\sin^2\frac{2\pi m}{M}+\rho \sin^2
\frac{\pi (n+1/2)}{N}}\nonumber
\end{eqnarray}
The sum over m in the Eq. (\ref{cyl3}) can be carried out using
the identities given by Eq. (\ref{a5}) and (\ref{a51}). This
yields
\begin{eqnarray}
R_{M\times N}^{cyl}(r,s)& =&\frac{\sqrt{r s}}{2N} \sum_{n=0}^{
N-1}\frac{\cos^2{\frac{\pi n}{N}}}{\sin \frac{2 \pi
n}{N}\sqrt{1+\rho\sin^2{\frac{2 \pi n}{N}}}}{\rm
coth}\left[\frac{M}{2}\,\omega\left(\frac{2\pi n}{N}\right)+i
\pi/2\right]\label{cyl21}\\
&+&\frac{\sqrt{r s}}{2N} \sum_{n=0}^{ N-1}\frac{\cos^2{\frac{\pi
(n+1/2)}{N}}}{\sin \frac{2 \pi
(n+1/2)}{N}\sqrt{1+\rho\sin^2{\frac{2 \pi (n+1/2)}{N}}}}{\rm
coth}\frac{M}{2}\,\omega\left(\frac{2\pi
(n+1/2)}{N}\right).\nonumber
\end{eqnarray}
Introducing function $f(x)$ given by Eq. (\ref{fxcyl}) we finally
arrived to the Eq. (\ref{cylRfin}).

\section{Asymptotic expansion of $R_{M \times N}(r,s)$}
In Sec. II we have shown that the  resistance between two maximum
separated nodes on an $M \times N$ rectangular network of
resistors with various boundary conditions can be expressed, in
terms of the
 function  $G_{1/2,0}(\rho,M,N)$ and $G_{0,1/2}(\rho,M,N)$ only,
 (see
Eqs.(\ref{2dRfin}), (\ref{RRfin}) and (\ref{cylRfin})).

Based on such results, one can use the method proposed by
Ivashkevich, Izmailian, and Hu \cite{izmailian2002} to derive the
asymptotic expansion of the $G_{\alpha,\beta}(\rho,M,N)$ in terms
so-called Kronecker's double series \cite{Weil}, which are
directly related to elliptic $\theta$ functions (see Appendix
\ref{Asymptotic}).

After reaching this point, one can easily write down all the terms
of the exact asymptotic expansion for the resistance between two
maximum separated nodes of the network ($R_{M \times N}(r,s)$)
using Eq. (\ref{ExpansionGfin}). We have found that the exact
asymptotic expansion of the $R_{M \times N}(r,s)$ for free,
periodic and cylindrical boundary conditions can be written as Eq.
(\ref{RmnAsymptotic}).

(i) For the free boundary conditions we obtain
\begin{eqnarray}
\frac{1}{s}R_{M \times N}^{free}(r,s) &=&\frac{2
\sqrt{\rho}}{\pi}\left[2\ln{N}+
\int_{0}^{\pi}\!\!\varphi(x)~\!{\rm d}x+ 2C_E +
4\ln{2}-\frac{\pi\sqrt{\rho}}{2}-2 \ln{\theta_2(i\sqrt{\rho}\,
\xi)\theta_4(i\sqrt{\rho}\, \xi)} \right]
\nonumber\\
&-& \frac{\sqrt{\rho}}{\pi} \sum_{p=1}^{\infty}\left(\frac{\pi^2
\xi}{S}\right)^{p}\frac{\Omega_{2p}}{p(2p)!} \left[{\rm
K}_{2p}^{0,1/2}(i\sqrt{\rho}\,\xi)+{\rm
K}_{2p}^{1/2,0}(i\sqrt{\rho}\,\xi)\right]
\label{ExpansionGfinfree}
\end{eqnarray}
where integral $\int_{0}^{\pi}\!\!\varphi(x)~\!{\rm d}x$ is given
by Eq. (\ref{intfree}).

Thus, the coefficients $c_{2p}(\rho,\xi)$ (p=1,2,..) in the
expansion Eq. (\ref{RmnAsymptotic}) explicitly given by
\begin{equation}
c_{2p}(\rho,\xi)=\frac{\pi^{2p-1} \xi^p\sqrt{\rho}}{p(2p)!}
\Omega_{2p} \left[{\rm K}_{2p}^{0,1/2}(i\sqrt{\rho}\,\xi)+{\rm
K}_{2p}^{1/2,0}(i\sqrt{\rho}\,\xi)\right] \label{c2p}
\end{equation}
where the differential operators $\Omega_{2p}$ is given by Eq. (
\ref{Omega2p}) and $K_{2p}^{0,1/2}(i\sqrt{\rho}\xi)$,
$K_{2p}^{1/2,0}(i\sqrt{\rho}\xi)$ are the Kronecker's double
series which can all be expressed in terms of the elliptic
$\theta_k(i\sqrt{\rho}\xi)$ ($k = 2, 3, 4$) functions only (see
Appendix \ref{KroneckerDoubleSeries}).

Here we list the first few coefficients in the expansion given by
Eq. (\ref{RmnAsymptotic})
\begin{eqnarray}
c(\rho)&=&\frac{2\sqrt{\rho}}{\pi}\label{cfree}\\
c_0(\rho,\xi)&=&\frac{2\sqrt{\rho}}{\pi}\left(2\ln{\frac{8}{\pi}}+2
C_E-1-\ln{\xi(1+\rho) -\frac{\pi
\sqrt{\rho}}{2}+\frac{\rho-1}{\sqrt{\rho}}}{\rm
arctan}\sqrt{\rho}-2\ln{\theta_2\theta_4}\right)
\label{c0free}\\
c_2(\rho,\xi)&=& \frac{\pi \tau_0
}{72}\left(5-3\rho+(1+\rho)\tau_0\frac{\partial}{\partial\tau_0}\right)
(\theta_4^4-\theta_2^4)
\label{c2free}\\
&=& \frac{\pi\tau_0}{72}\left((5-3\rho)(\theta_4^4-\theta_2^4)+\pi
\tau_0 (1+\rho)\theta_3^4\theta_4^4+4\tau_0
(1+\rho)(\theta_4^4-\theta_2^4)\frac{\partial}{\partial\tau_0}\ln
{\theta_2}\right) \nonumber\\
~&\vdots&~ \nonumber
\end{eqnarray}
To simplify the notation we have use the short hand
\begin{equation}
\theta_k=\theta_k(i \tau_0), \qquad k = 2, 3, 4,
\label{shorthand2}
\end{equation}
where $\tau_0=\xi\sqrt{\rho}$.

We have also used the following relations between derivatives of
the elliptic functions
$$
\frac{\partial}{\partial \tau_0}\ln{{\theta}_3} =
\frac{\pi}{4}{\theta}_4^4+\frac{\partial}{\partial
\tau_0}\ln{{\theta}_2} \qquad \mbox{and} \qquad
\frac{\partial}{\partial \tau_0}\ln{{\theta}_4} =
\frac{\pi}{4}{\theta}_3^4+\frac{\partial}{\partial
\tau_0}\ln{{\theta}_2}
$$
Note that elliptic functions ${\theta}_2, {\theta}_3, {\theta}_4$
can be expressed through the complete elliptic integral of the
first kind $K=K(k)$ and second kind $E=E(k)$ as
\begin{equation}
{\theta}_2=\sqrt{\frac{2 k K(k)}{\pi}}, \qquad
{\theta}_3=\sqrt{\frac{2 K(k)}{\pi}}, \qquad
{\theta}_4=\sqrt{\frac{2 k' K(k)}{\pi}} \label{K28}
\end{equation}
where
\begin{eqnarray}
K(k)&=& \int_0^{\pi/2}\frac{{\rm d}x}{\sqrt{1-k^2\sin^2{x}}},\label{EllipticIntK}\\
 E(k) &=& \int_0^{\pi/2}\sqrt{1-k^2\sin^2{x}}~\!{\rm d}x.
\label{EllipticIntE}
\end{eqnarray}

With the help of the identities
$$
\frac{\partial}{\partial \tau_0}\ln{\theta_2} =
-\frac{1}{2}{\theta}_3^2 E, \qquad \mbox{and} \qquad
\frac{\partial E}{\partial \tau_0}
=\frac{\pi^2}{4}{\theta}_3^2{\theta}_4^4-\frac{\pi}{2}{\theta}_4^4
E
$$
one can express all derivatives of the elliptic functions in terms
of the elliptic functions ${\theta}_2, {\theta}_3, {\theta}_4$ and
the complete elliptic integral of the second kind $E=E(k)$.

For the case $M=N$ and $r=s$ ($\xi=1, \rho=1$) we reproduced the
result of Essam and Wu \cite{EssamWu2009}:
\begin{equation}
\frac{1}{s}R_{N \times
N}^{free}(s,s)=\frac{4}{\pi}\ln{N}+c_0+\frac{c_2}{N^2}+...
\label{wu}
\end{equation}
with $c_0=0.07731889390945876...$ and
$c_2=0.26607044163847837...$.

(ii) For the periodic boundary conditions we obtain
\begin{eqnarray}
\frac{1}{s}R_{M \times N}^{tor}(r,s) &=&\frac{
\sqrt{\rho}}{2\pi}\left[2\ln{\frac{N}{2}}+
\int_{0}^{\pi}\!\!\varphi(x)~\!{\rm d}x+ 2C_E + 4\ln{2}-2
\ln{\theta_2(i\sqrt{\rho}\, \xi)\theta_4(i\sqrt{\rho}\, \xi)}
\right]
\nonumber\\
&-& \frac{\sqrt{\rho}}{4\pi}
\sum_{p=1}^{\infty}\frac{\Omega_{2p}}{p(2p)!} \left(\frac{4 \pi^2
\xi}{S}\right)^{p}\left[{\rm
K}_{2p}^{0,1/2}(i\sqrt{\rho}\,\xi)+{\rm
K}_{2p}^{1/2,0}(i\sqrt{\rho}\,\xi)\right] \label{ExpansionGfintor}
\end{eqnarray}
where integral $\int_{0}^{\pi}\!\!\varphi(x)~\!{\rm d}x$  is given
by Eq. (\ref{inttor}).

The first few coefficients in the expansion are
\begin{eqnarray}
c(\rho)&=&\frac{\sqrt{\rho}}{2\pi}\label{ctor}\\
c_0(\rho,\xi)&=&\frac{\sqrt{\rho}}{2\pi}\left(2\ln{\frac{4}{\pi}}+2
C_E-\ln{\xi(1+\rho)} -2\ln{\theta_2\theta_4}\right)
\label{c0tor}\\
c_2(\rho,\xi)&=&\frac{\pi \tau_0
}{72}\left(3\rho-1+(1+\rho)\tau_0\frac{\partial}{\partial\tau_0}\right)
(\theta_4^4-\theta_2^4)
\label{c2tor}\\
&=& \frac{\pi\tau_0}{72}\left((3\rho-1)(\theta_4^4-\theta_2^4)+\pi
\tau_0 (1+\rho)\theta_3^4\theta_4^4+4\tau_0
(1+\rho)(\theta_4^4-\theta_2^4)\frac{\partial}{\partial\tau_0}\ln
{\theta_2}\right) \nonumber\\~&\vdots&~\nonumber
\end{eqnarray}

For the case $M=N$ and $r=s$ ($\xi=1, \rho=1$) we obtain:
\begin{equation}
\frac{1}{s}R_{N \times
N}^{tor}(s,s)=\frac{1}{\pi}\ln{N}+c_0+\frac{c_2}{N^2}+...
\label{wuper}
\end{equation}
with $c_0=0.2078490664166084...$ and $c_2=0.2660704416384784...$.

(iii) And for the cylindrical boundary conditions
\begin{eqnarray}
\frac{1}{s}R_{M \times N}^{cyl}(r,s) &=&\frac{
\sqrt{\rho}}{\pi}\left[2\ln{N}+
\int_{0}^{\pi}\!\!\varphi(x)~\!{\rm d}x+ 2C_E + 4\ln{2}-2
\ln{\theta_2(i\sqrt{\rho}\, \xi/2)\theta_4(i\sqrt{\rho}\, \xi/2)}
\right]
\nonumber\\
&-& \frac{\sqrt{\rho}}{2\pi}
\sum_{p=2}^{\infty}\frac{\Omega_{2p}}{p(2p)!} \left(\frac{\pi^2
\xi}{S}\right)^{p}\left[{\rm
K}_{2p}^{0,1/2}(i\sqrt{\rho}\,\xi/2)+{\rm
K}_{2p}^{1/2,0}(i\sqrt{\rho}\,\xi/2)\right]
\label{ExpansionGfincyl}
\end{eqnarray}
where integral $\int_{0}^{\pi}\!\!\varphi(x)~\!{\rm d}x$  is given
by Eq. (\ref{intcyl}).

The first few coefficients in the exact asymptotic expansion  are

\begin{eqnarray}
c(\rho)&=&\frac{\sqrt{\rho}}{\pi}\label{ccyl}\\
c_0(\rho,\xi)&=&\frac{\sqrt{\rho}}{\pi}\left(2\ln{\frac{8}{\pi}}+2
C_E-\ln{\xi(1+\rho)-\frac{2}{\sqrt{\rho}}}{\rm
arctan}\sqrt{\rho}-2\ln{\theta_2\theta_4}\right)
\label{c0cyl}\\
c_2(\rho,\xi)&=&\frac{\pi \tau_1
}{72}\left(3\rho+5+(1+\rho)\tau_1\frac{\partial}{\partial\tau_1}\right)
(\theta_4^4-\theta_2^4)
\label{c2cyl}\\
&=& \frac{\pi\tau_1}{72}\left((5+3\rho)(\theta_4^4-\theta_2^4)+\pi
\tau_1 (1+\rho)\theta_3^4\theta_4^4+4\tau_1
(1+\rho)(\theta_4^4-\theta_2^4)\frac{\partial}{\partial\tau_1}\ln
{\theta_2}\right) \nonumber\\~&\vdots&~\nonumber
\end{eqnarray}
Here we have use the short hand
\begin{equation}
\theta_k=\theta_k(i \tau_1), \qquad k = 2, 3, 4,
\label{shorthand2a}
\end{equation}
with $\tau_1=\xi\sqrt{\rho}/2$.

For the case $M=N$ and $r=s$ ($\xi=1, \rho=1$) we obtain:
\begin{equation}
\frac{1}{s}R_{N \times
N}^{cyl}(s,s)=\frac{2}{\pi}\ln{N}+c_0+\frac{c_2}{N^2}+...
\label{wucyl}
\end{equation}
with $c_0=0.36172475911729557...$ and
$c_2=-0.28448262410676656...$.

Let us now consider the behavior of the coefficients
$c_{2k}(\rho,\xi)$  in the asymptotic expansion of the
 resistance between two maximum separated nodes on the
rectangular network under the Jacobi transformation (see Appendix
\ref{jacobiTransformation}). Using Eq. (\ref{def preob}) and Eq.
(\ref{def_Jacobi_transformation2Kron}) we can easily check that
$c_{2k}(\rho,\xi)$ (for all k) are invariant under transformation
$\tau_0 \to 1/\tau_0$, where
\begin{eqnarray}
\tau_{0}&=&\xi \; \sqrt{\rho}
\qquad \mbox{for free and periodic b.c.} \label{effective1}\\
\tau_{0}&=&\xi \; \sqrt{\rho}/2 \qquad \mbox{for cylindrical b.c.}
\label{effectivecyl1}
\end{eqnarray}
Using the properties of the $\theta$ -functions   and of the
functions $K^{\alpha,\beta}_{2p}$ (see Eq. (\ref{def preob}) and
(\ref{def_Jacobi_transformation2Kron})) we can easily check from
Eq. (\ref{ExpansionGfin}) that the $\ln
G_{\alpha,\beta}(\rho,M,N)$ have the following behavior under the
transformation $\tau_0 \to 1/\tau_0$:
\begin{eqnarray}
G_{1/2,0}(\rho,M,N) \to G_{0,1/2}(\rho,M,N) \label{Zinvariant}
\end{eqnarray}

Equations (\ref{2dRfin}), (\ref{RRfin}), (\ref{cylRfin}) and
(\ref{Zinvariant}) imply that the resistance $R_{\{M\times
N\}}(r,s)$ between two maximum separated nodes of the network for
all above mentioned boundary conditions is invariant under
transformation $\tau_0 \to 1/\tau_0$. This actually means that
$\xi_{eff}$ given by Eqs. (\ref{effective}) and
(\ref{effectivecyl}) can be regarded as the effective aspect
ratio.

\section{Discussion}
\label{last part}

In Fig. 2 we plot the conventional aspect-ratio ($\xi$) dependence
of the finite-size correction term $c_0(\xi)$ for the resistance
between two maximum separated nodes on a $M \times N$ resistor
network with the free (solid lines), toroidal (dashed lines) and
cylindrical (dot-dashed lines) boundary conditions for several
values of the factor $\rho$: (a) for $\rho = 1$ and (b) for
$\rho=4$. We use the logarithmic scales in the horizontal axis.
The finite-size correction term $c_0(\xi)$ at first decrease until
$\xi=\xi_{min}$: $\xi_{min}=1$ for $\rho = 1$ and  $\xi_{min}=1/2$
 for $\rho = 4$. Note that $\xi_{min}=1/\sqrt{\rho}$ for arbitrary value of $\rho$.
With further increase of $\xi$ it reverses directions, increasing
monotonically to infinity. For large enough $\xi$ ($\xi \gg 1$),
the finite size properties of the resistor network with
cylindrical boundary condition and those of the torus become the
same, which means that the boundaries along the shorter direction
determine the finite size properties of the system; for both
cylindrical boundary condition and the torus, the boundary
condition along the shorter direction is the periodic one. For
small enough $\xi$ ($\xi \ll 1$), the finite size properties of
the resistor network with free boundary condition and those of the
cylinder become the same because the boundaries along the shorter
directions for these two are the same, that is, the free boundary
condition.
\begin{figure}
\epsfxsize=70mm \vbox to2.5in{\rule{0pt}{2.5in}}
\includegraphics{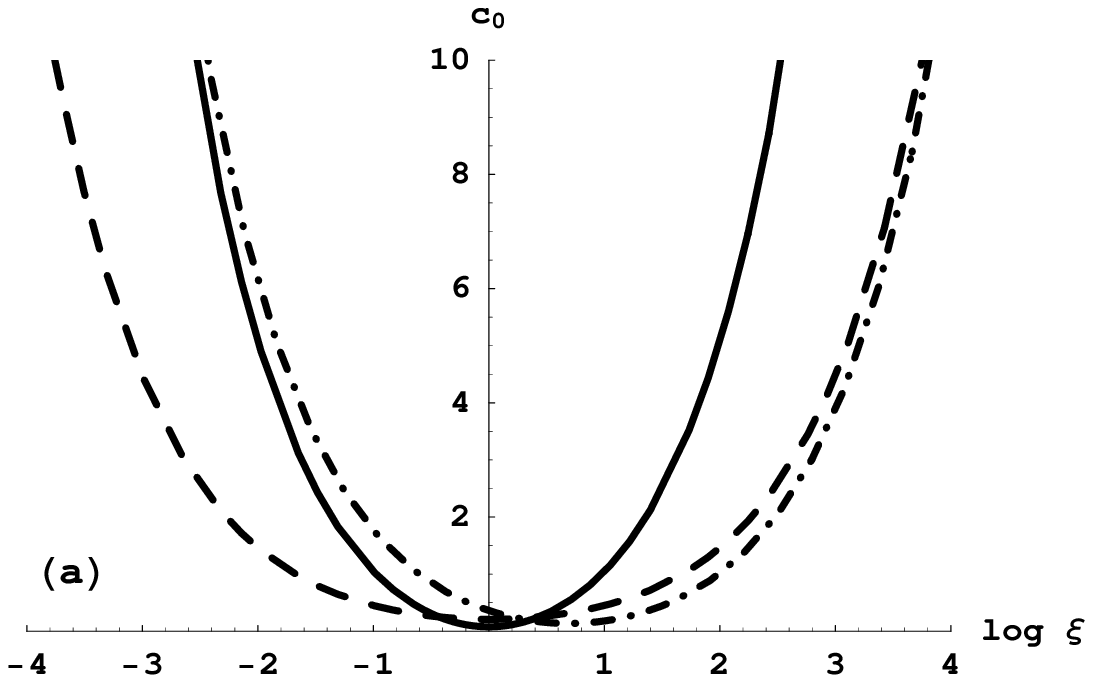} \includegraphics{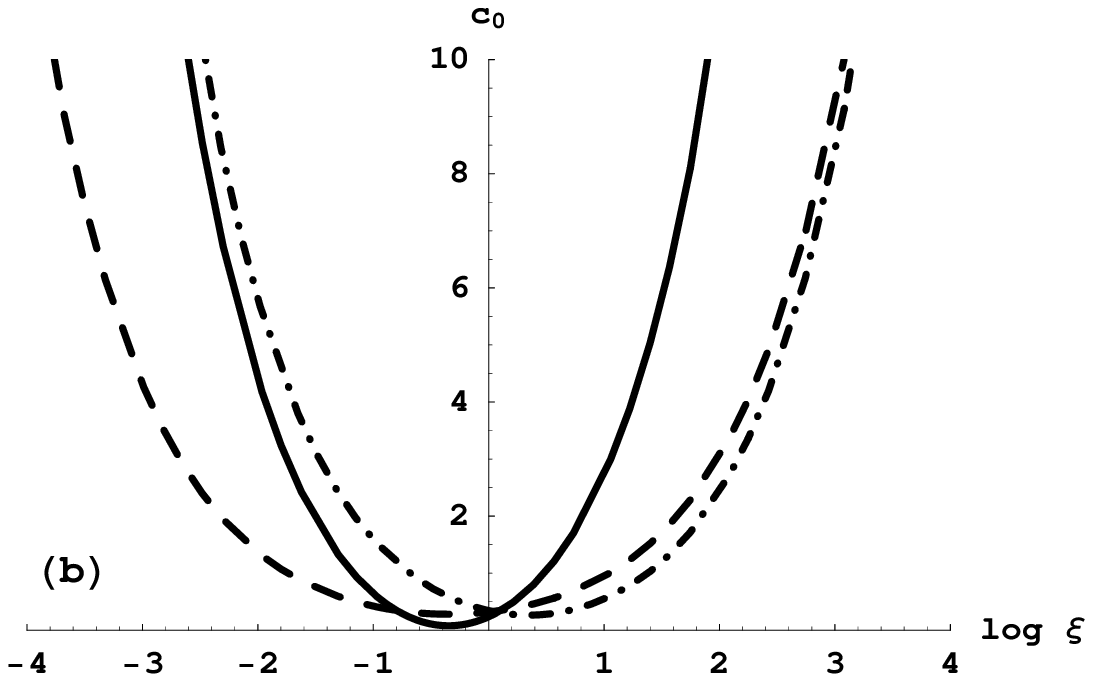} \caption{Conventional
aspect-ratio ($\xi$) dependence of finite-size correction term
$c_0$ for the resistance between two maximum separated nodes on a
$M \times N$ resistor network with the free (solid lines),
toroidal (dashed lines) and cylindrical (dot-dashed lines)
boundary conditions: (a) for $\rho = 1$ and (b) for $\rho = 4$. We
use the natural logarithmic scales for the horizontal axis.}
\end{figure}

In Fig. 3 we plot the effective aspect-ratio ($\xi_{eff}$)
dependence of the finite-size correction term $c_0$ for the
resistance between two maximum separated nodes on a $M \times N$
resistor network with the free (solid lines), toroidal (dashed
lines) and cylindrical (dot-dashed lines) boundary conditions for
several values of the factor $\rho$: (a) for $\rho = 1$ and (b)
for $\rho=4$. We use the logarithmic scales in the horizontal
axis. We can see that finite size correction terms $c_0$ are
invariant under transformation $\xi_{eff} \to 1/\xi_{eff}$. The
finite-size correction term $c_0(\xi)$ at first decrease until
$\xi_{eff}=1$ for all boundary conditions and for for arbitrary
value of $\rho$. With further increase of $\xi_{eff}$ it reverses
directions, increasing monotonically to infinity.
\begin{figure}
\epsfxsize=70mm \vbox to2.5in{\rule{0pt}{2.5in}}
\includegraphics{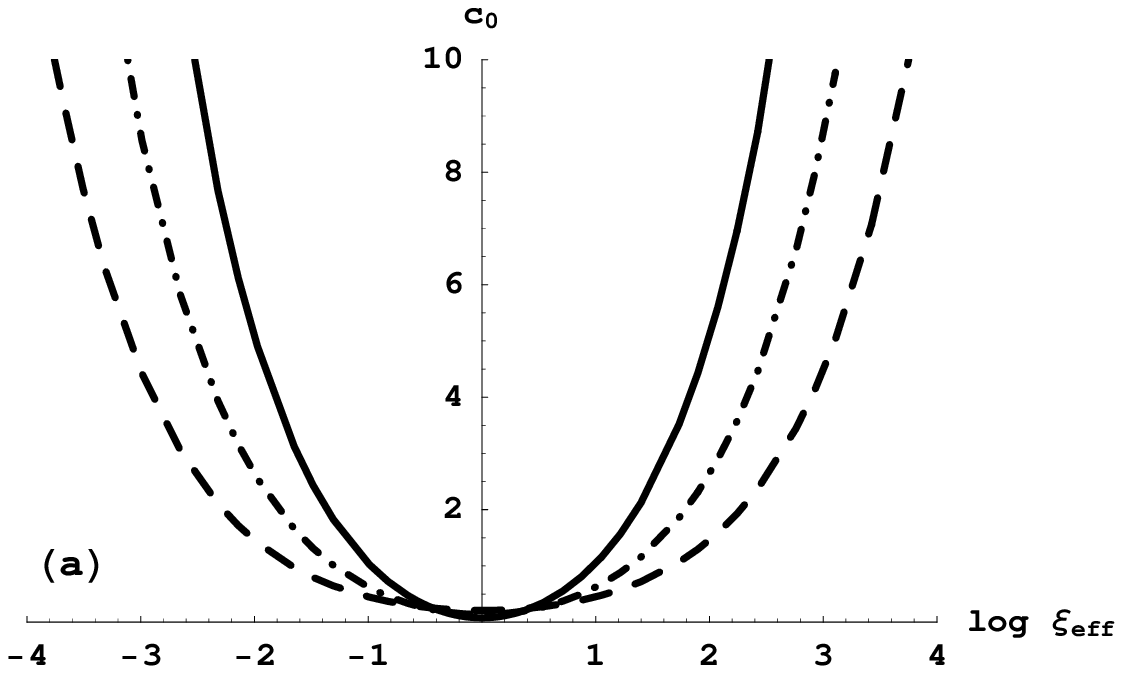} \includegraphics{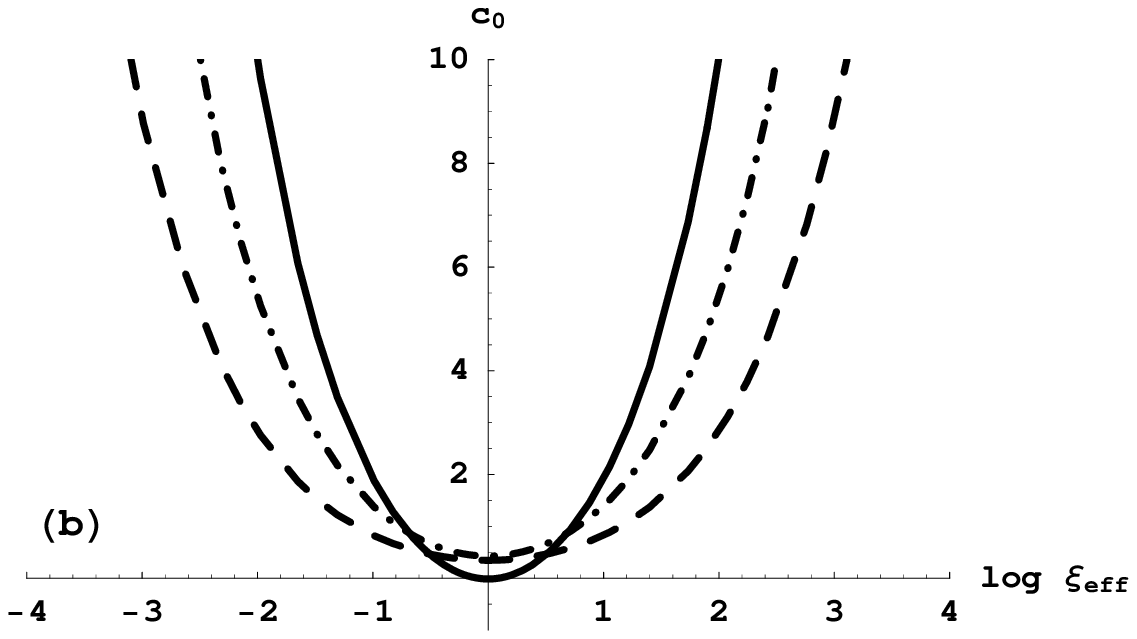} \caption{Effective
aspect-ratio ($\xi_{eff}$) dependence of finite-size correction
term $c_0$ for the resistance between two maximum separated nodes
on a $M \times N$ resistor network with the free (solid lines),
toroidal (dashed lines) and cylindrical (dot-dashed lines)
boundary conditions: (a) for $\rho = 1$ and (b) for $\rho = 4$.
 We use the natural logarithmic scales for
the horizontal axis.}
\end{figure}
In Fig. 4 we plot the $\rho$ dependence of the finite-size
correction term $c_0(\rho)$ for the resistance between two maximum
separated nodes on a $M \times N$ resistor network with the free
(solid lines), toroidal (dashed lines) and cylindrical (dot-dashed
lines) boundary conditions  for several values of the aspect-ratio
$\xi$: (a) for $\xi = 1$ and (b) for $\xi=4$.  The finite-size
correction term $c_0(\rho)$ at first decrease until
$\rho=\rho_{min}$. Note that value of $\rho_{min}$ depends on the
boundary conditions as well on the value of the aspect ratio
$\xi$.
With further increase of $\rho$ it reverses directions, increasing
monotonically to infinity.

\begin{figure}
\epsfxsize=70mm \vbox to2.5in{\rule{0pt}{2.5in}}
\includegraphics{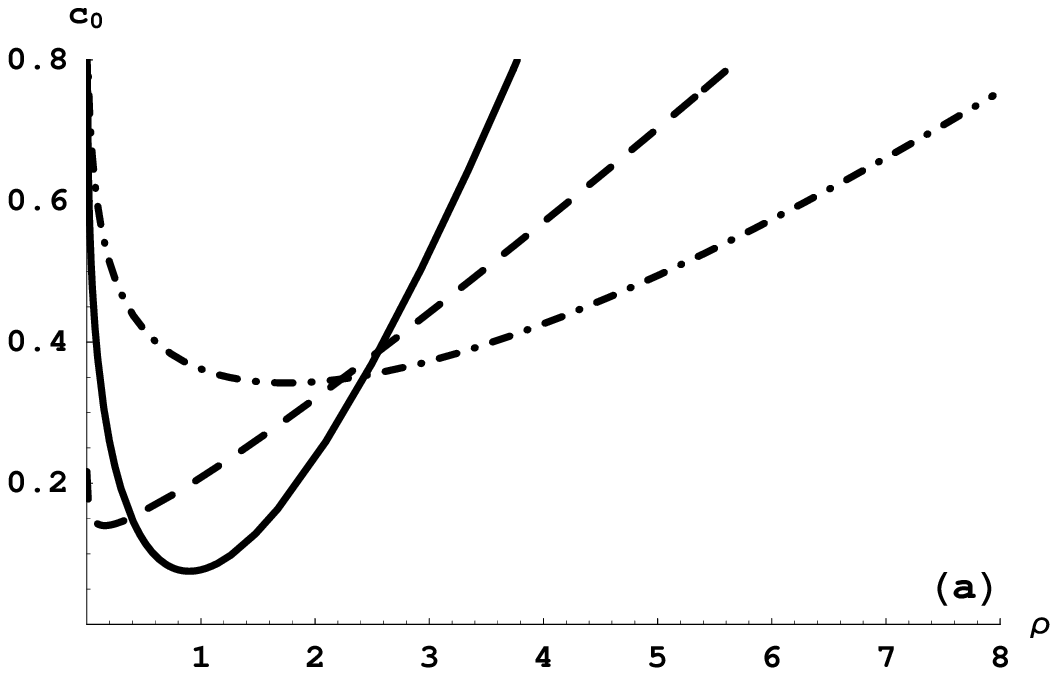} \includegraphics{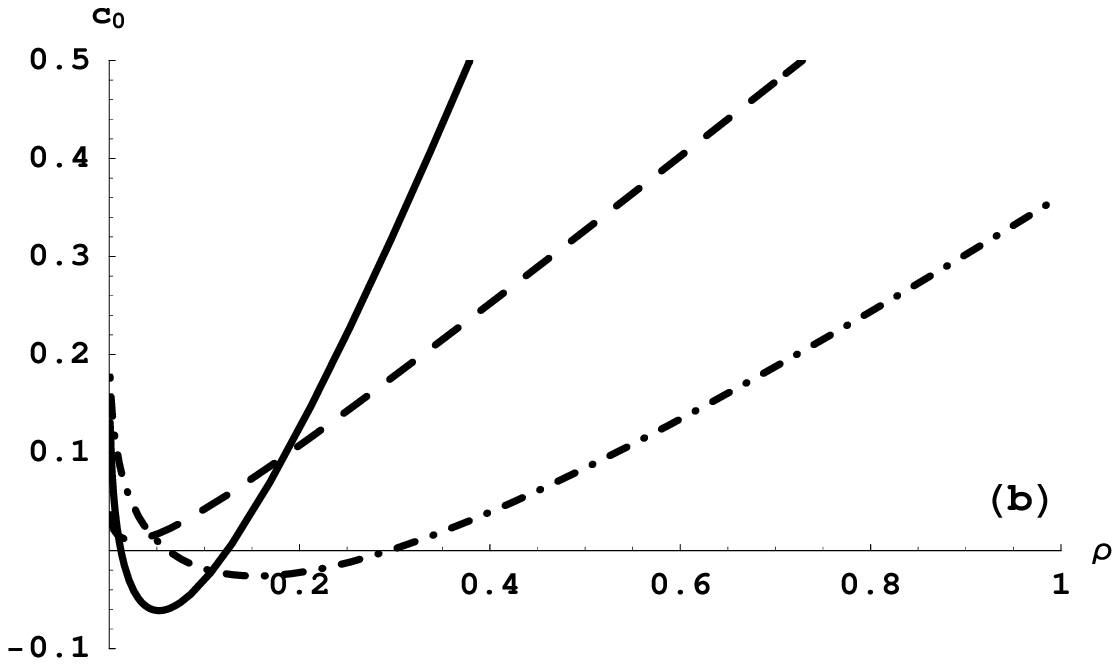}   \caption{The ($\rho$)
dependence of finite-size correction term $c_0$ for the resistance
between two maximum separated nodes on a $M \times N$ resistor
network with the free (solid lines), toroidal (dashed lines) and
cylindrical (dot-dashed lines) boundary conditions: (a) for $\xi =
1$ and (b) for $\xi = 4$.}
\end{figure}

In the present paper, we study the two-point resistor problem on
planar $M \times N$ rectangular lattices with free, periodic and
cylindrical boundary conditions. Using the exact expression for
the resistance between arbitrary two nodes for finite rectangular
network obtained in \cite{Wu2004} and the IIH¡¦s algorithm
\cite{izmailian2002}, we derive the exact asymptotic expansion of
the corner-to-corner resistance on the rectangular network for all
above mentioned boundary conditions. All corrections to scaling
are analytic.

\section{Acknowledgments}

This work was partially supported by the National Science Council
of Republic of China (Taiwan) under Grant No. NSC
96-2112-M-033-006. N.Sh.I is supported in part by National Center
for Theoretical Sciences: Physics Division, National Taiwan
University, Taipei, Taiwan.

\appendix

\section{Asymptotic expansion of $G_{\alpha,\beta}(\rho,{\cal M},{\cal N})$}
\label{Asymptotic}

Using the expansion of the ${\rm coth} x$
$$
{\rm coth}\, x=1+2\sum_{m=1}^{\infty}e^{-2 m x }
$$
we can transform the Eq. (\ref{Gab}) in the following form
\begin{eqnarray}
G_{\alpha,\beta}(\rho,{\cal M},{\cal N})&=&~{\tt Re}{\cal M}
\sum_{n=0}^{{\cal N}-1}
f\!\left(\textstyle{\frac{\pi(n+\alpha)}{{\cal N}}}\right) {\rm
coth}\left[{\cal M} f\!\left(\textstyle{\frac{\pi(n+\alpha)}{{\cal
N}}}\right) +i\pi\beta\right]
\label{ExpansionG} \\
&=&{\cal M} \sum_{n=0}^{{\cal
N}-1}f\!\left(\textstyle{\frac{\pi(n+\alpha)}{{\cal N}}}
\right)+2~{\tt Re} {\cal M} \sum_{n=0}^{{\cal
N}-1}\sum_{m=1}^{\infty}
f\!\left(\textstyle{\frac{\pi(n+\alpha)}{{\cal N}}}\right)
e^{-2m\left[{\cal M}\omega\left(\frac{\pi(n + \alpha)}{{\cal
N}}\right) + i\pi\beta\right]} \nonumber
\end{eqnarray}
where $f(x)$ is given by Eqs. (\ref{fxfree}), (\ref{fxtor}) and
(\ref{fxcyl}) for free, periodic and cylindrical b.c.
respectively.

Using Taylor's theorem, the asymptotic expansion of the $f(x)$ for
all boundary conditions can be written in the following form
\begin{equation}
f(x) = \frac{1}{x}\left[1+\sum_{p=1}^{\infty}\frac
{\kappa_{2p}}{(2p)!}x^{2p} \right]\label{omegaTaylor1}
\end{equation}
where $\kappa_{2}=\rho-5/3$, $\kappa_{4}=-3\rho^2-14\rho+67/15$,
etc for free boundary conditions $\kappa_{2}=-\rho+1/3$,
$\kappa_{4}=9\rho^2+2\rho+7/15$, etc. for periodic boundary
conditions and $\kappa_{2}=-\rho-5/3$,
$\kappa_{4}=9\rho^2+14\rho+67/15$, etc. for cylindrical boundary
conditions.

Note that function $f(x)$ can be represent as
\begin{equation}
f(x)=\frac{1}{x}\;\exp\left\{
{\sum_{p=1}^{\infty}\frac{\varepsilon_{2p}}{(2p)!}}x^{2p} \right\}
\label{owega''}
\end{equation}
where coefficients $\varepsilon_{2p}$ and $\kappa_{2p}$ are
related to each other through relation between moments and
cumulants (Appendix \ref{MomentsCumulants})
\begin{eqnarray}
\kappa_{2}&=&\varepsilon_2\nonumber\\
\kappa_{4}&=&\varepsilon_4+3\varepsilon_2^2\,
\nonumber\\
&\vdots& \nonumber
\end{eqnarray}

We will need also the Taylor expansion of the $\omega(x)$ given by
Eq. (\ref{omega})
\begin{equation}
\omega(x)=x\left(\lambda+\sum_{p=1}^{\infty}
\frac{\lambda_{2p}}{(2p)!}\;x^{2p}\right) \label{omegaTaylor}
\end{equation}
where $\lambda=\sqrt{\rho}$, $\lambda_2=-\sqrt{\rho} (1+\rho)/3$,
$\lambda_4=\sqrt{\rho} (1+10\rho+9\rho^2)/5$, etc. for all
boundary conditions.

In what follows, we shall not use the special values of these
coefficients assuming the possibility for generalizations.

The asymptotic expansion of $G_{\alpha,\beta}(\rho,{\cal M},{\cal
N})$ can be derived in the similar way as it has done in Ref.
\cite{izmailian2002} for the second derivative of the partition
function with twisted boundary condition (see Eq. (18) in Ref.
\cite{izmailian2002}). Following along the same line as in Ref.
\cite{izmailian2002} we can obtain for the asymptotic expansion of
$G_{\alpha,\beta}(\rho,{\cal M},{\cal N})$ the following
expression
\begin{eqnarray}
G_{\alpha,\beta}(\rho,{\cal M},{\cal N}) &=&\frac{2 {\cal
S}}{\pi}\left[\ln{\cal N}+\frac{1}{2}
\int_{0}^{\pi}\!\!\varphi(x)~\!{\rm d}x+ C_E + 2\ln{2}-2
\ln{\left|\theta_{\alpha,\beta}(i\lambda \xi)\right|} \right]
\nonumber\\
&+&\frac{1}{2}
\left(\kappa_2\frac{\partial}{\partial\lambda}+\lambda_2
\frac{\partial^2}{\partial\lambda^2}\right)
\ln\left|\frac{\theta_{\alpha,\beta}(i\lambda \xi)}{\eta(i\lambda
\xi)}\right|
\label{ExpansionGfin}\\
&-& \pi \xi \sum_{p=2}^{\infty}\left(\frac{\pi^2 \xi}{{\cal
S}}\right)^{p-1}\frac{\Omega_{2p}}{p(2p)!} {\tt Re}\; {\rm
K}_{2p}^{\alpha,\beta}(i\lambda\xi) \nonumber
\end{eqnarray}
where ${\cal S} = {\cal M} {\cal N}$, $\xi = {\cal M}/{\cal N}$,
$C_E$ is the Euler constant, $\eta(\tau)$ is the Dedekind - $\eta$
function
\begin{equation}
\eta(\tau)=e^{\pi i \tau /12}\prod_{n=1}^\infty\left[ 1-e^{2\pi
i\tau n}\right],
\end{equation}
$K_{2p+2}^{\alpha,\beta}(\tau)$ is Kronecker's double series
(Appendix \ref{KroneckerDoubleSeries}) and
 $\theta_{\alpha,\beta}(\tau)$ is elliptic theta function (Appendix
 \ref{ThetaFunctions}).

The differential operators $\Omega_{2p}$ that have appeared here
can be expressed via coefficients
$\omega_{2p}=\varepsilon_{2p}+\lambda_{2p}
\frac{\partial}{\partial\lambda}$ as
\begin{eqnarray}
{\Omega}_{2}&=&\omega_2\nonumber\\
{\Omega}_{4}&=&\omega_4+3\omega_2^2\, \label{Omega2p}\\ &\vdots&
\nonumber
\end{eqnarray}

The function $\varphi(x)$ is defined as
\begin{equation}
\varphi(x)=f(x)-\frac{1}{x}-\frac{1}{\pi-x} \label{varphi}
\end{equation}

Thus, the $\int_{0}^{\pi}\!\!\varphi(x)~\!{\rm d}x$ depend on the
boundary conditions and given by
\begin{eqnarray}
\int_{0}^{\pi}\!\!\varphi(x)~\!{\rm
d}x&=&-1+2\ln\frac{2}{\pi}-\ln(1+\rho)+\frac{\rho-1}{\sqrt{\rho}}{\rm
arctan}\sqrt{\rho} \qquad \mbox{for free b.c.,}\label{intfree}\\
\int_{0}^{\pi}\!\!\varphi(x)~\!{\rm d}x&=&
2\ln\frac{2}{\pi}-\ln(1+\rho) \hspace{5cm} \mbox{for periodic
b.c.,}\label{inttor}\\
\int_{0}^{\pi}\!\!\varphi(x)~\!{\rm
d}x&=&2\ln\frac{2}{\pi}-\ln(1+\rho)-\frac{2}{\sqrt{\rho}}{\rm
arctan}\sqrt{\rho} \qquad \qquad \mbox{for cylindrical
b.c..}\label{intcyl}
\end{eqnarray}

We are interested in the asymptotic expansion of
$G_{\alpha,\beta}(\rho,{\cal M},{\cal N})$ with
$(\alpha,\beta)=(0,\frac{1}{2})$ and $(\frac{1}{2},0)$.

\section{Relation between moments and cumulants}
\label{MomentsCumulants}

Moments $Z_k$ and cumulants $F_k$ which enters the expansion of
exponent
$$\exp\left\{\;\sum_{k=1}^{\infty}\frac{x^k}{k!}\,F_k\,\right\}
=1+\sum_{k=1}^{\infty}\frac{x^k}{k!}\,Z_k$$
are related to each other as \cite{Prohorov}
\begin{eqnarray}
Z_1&=&F_1\nonumber\\ Z_2&=&F_2+F^2_1\nonumber\\
Z_3&=&F_3+3F_1F_2+F^3_1\nonumber\\
Z_4&=&F_4+4F_1F_3+3F_2^2+6F_1^2F_2+F^4_1\nonumber\\
~&\vdots&~\nonumber\\ Z_k&=&\sum_{r=1}^{k}\sum
\left(\frac{F_{k_1}}{k_1!}\right)^{i_1}\ldots
\left(\frac{F_{k_r}}{k_r!}\right)^{i_r} \frac{k!}{i_1!\ldots
i_r!}\nonumber
\end{eqnarray}
where summation is over all positive numbers $\{i_1\ldots i_r\}$
and different positive numbers $\{k_1,\ldots,k_r\}$ such that $k_1
i_1+\ldots+ k_r i_r=k$.

\section{Elliptic Theta Functions}
\label{ThetaFunctions} In this appendix we gather all the
definitions and properties of the Jacobi's $\theta$-functions and
Kronecker's double series needed in this paper. We adopt the
following definition of the elliptic $\theta$-functions:
\begin{eqnarray}
\theta_{\alpha,\beta}(z,\tau)&=&\sum_{n\in Z} \exp\left\{ \pi
i\tau \left(n+\textstyle{\frac{1}{2}}-\alpha\right)^2+2\pi i
\left(n+\textstyle{\frac{1}{2}}-\alpha\right)\left(z-\textstyle{\frac{1}{2}}+\beta\right)
\right\}~~~~\nonumber\\ &=&\eta(\tau)\,\exp\left\{\textstyle{\pi
i\tau B_2^{\alpha}+2\pi
i\big(\frac{1}{2}-\alpha\big)\big(z-\frac{1}{2}+\beta\big)}\right\}\nonumber\\
&\times&\prod_{n=0}^{\infty}\!\Big[\,1-e^{2\pi
i\tau\left(n+\alpha\right)-2\pi i\left(z+\beta\right)}\,\Big]
\Big[\,1-e^{2\pi i\tau\left(n+1-\alpha\right)+2\pi
i\left(z+\beta\right)}\,\Big]\label{definitionTheta}
\end{eqnarray}
where $B_2^{\alpha}$ is the Bernoulli polynomial $B_2^{\alpha} =
\alpha^2 - \alpha + \frac{1}{ 6}$ and $\eta(\tau)$ is Dedekind
$\eta$-function:
\begin{equation}
\eta(\tau)=e^{\pi i\tau B_2/2}\prod_{n=1}^{\infty}\Big[\,1-e^{2\pi
i \tau n}\,\Big]. \label{definitionEta}
\end{equation}
where $B_2 \equiv B_2^{0} = 1/6$.

The elliptic $\theta$-functions satisfies the heat equation
\begin{equation}
\frac{\partial}{\partial \tau}\theta_{\alpha,\beta}(z,\tau) =
\frac{1}{4\pi i}\frac{\partial^2}{\partial
z^2}\theta_{\alpha,\beta}(z,\tau) \label{heat}
\end{equation}
The relation of the functions $\theta_{\alpha,\beta}(z,\tau)$ with
the usual $\theta$-functions $\theta_i(z,\tau)$ $i=1,\ldots,4$ is
the following
\begin{eqnarray}
\theta_{0,0}(z,\tau)&=&\theta_1(z,\tau)\nonumber\\
\theta_{0,\frac{1}{2}}(z,\tau)&=&\theta_2(z,\tau)\nonumber\\
\theta_{\frac{1}{2},0}(z,\tau)&=&\theta_4(z,\tau)\nonumber\\
\theta_{\frac{1}{2},\frac{1}{2}}(z,\tau)&=&\theta_3(z,\tau)\nonumber
\end{eqnarray}
In this paper we will only need these functions evaluated at $z=0$
and $\tau = i \tau_0$ is a  pure imaginary aspect ratio. To
simplify the notation we will use the shorthand
\begin{eqnarray}
\theta_{\alpha,\beta}(i\tau_0)&=& \theta_{\alpha,\beta}(0,i\tau_0) \nonumber \\
\theta_i(i\tau_0) &=& \theta_i(0,i\tau_0) \nonumber
\end{eqnarray}

\section{Kronecker's Double Series}
\label{KroneckerDoubleSeries}

Kronecker's double series can be defined as \cite{Weil}
$${\rm K}_{p}^{\alpha,\beta}(\tau)= -\frac{p!}{(-2\pi i)^p}
\sum_{m,n\in Z \above0pt  (m,n)\neq(0,0)} \frac{e^{-2\pi
i(n\alpha+m\beta)}}{(n+\tau m)^{p}}$$
In this form, however, they cannot be directly applied to our
analysis. In \cite{izmailian2002} it was shown that Kronecker's
double series can be written as
\begin{equation}
{\rm K}_{p}^{\alpha,\beta}(\tau)={\rm B}_{p}^\alpha-p\sum_{m\neq
0}\sum_{n=0}^{\infty} (n+\alpha)^{p-1}~e^{2\pi im
(\tau(n+\alpha)-\beta)}\label{kroneker}
\end{equation}

The Kronecker's double series $K_{2p}^{\alpha, \beta}(\tau)$ for
the cases $(\alpha, \beta) = (0,0), (0,1/2), (1/2,0), (1/2,1/2)$
can all be expressed in terms of the elliptic $\theta(\tau)$
functions only.

Equations for $K_{2p}^{\alpha, \beta}(\tau)$ with $p = 2, 3, 4, 5$
and other useful relations for elliptic $\theta$-functions and
Kronecker's double series can be found in Refs.
\cite{izmailian2002,izmailian2002a,izmailian2003,izmailian2007}

Here we write down the Kronecker's double series $K_{2p}^{0,
1/2}(i\tau_0)$ and $K_{2p}^{1/2,0}(i\tau_0)$ that have appeared in
our asymptotic expansions
\begin{eqnarray}
 {\rm K}_{2}^{0,\frac{1}{2}}(i\tau_0)&=&
-\frac{1}{\pi}\frac{\partial}{\partial\tau_0}\ln\frac{\theta_2(i
\tau_0)}{\eta(i
\tau_0)}=\frac{\theta_3^4(i\tau_0)+\theta_4^4(i\tau_0)}{12},
\nonumber\\
{\rm K}_{2}^{\frac{1}{2},0}(i\tau_0)&=&
-\frac{1}{\pi}\frac{\partial}{\partial\tau_0}\ln\frac{\theta_4(i
\tau_0)}{\eta(i
\tau_0)}=-\frac{\theta_3^4(i\tau_0)+\theta_2^4(i\tau_0)}{12},
\label{listKroneker}\\
{\rm K}_{4}^{0,\frac{1}{2}}(i\tau_0)&=&\textstyle{\frac{1}{30}
[\frac{7}{8}\,\theta_2^8(i\tau_0)-\theta_3^4(i\tau_0)\theta_4^4(i\tau_0)]},\nonumber\\
{\rm K}_{4}^{\frac{1}{2},0}(i\tau_0)&=&\textstyle{\frac{1}{30}
[\frac{7}{8}\,\theta_4^8(i\tau_0)-\theta_2^4(i\tau_0)\theta_3^4(i\tau_0)]},\nonumber\\
{\rm K}_{6}^{0,\frac{1}{2}}(i\tau_0)&=&\textstyle{\frac{1}{84}
[\theta_3^4(i\tau_0)+\theta_4^4(i\tau_0)][\frac{31}{16}\,\theta_2^8(i\tau_0)+
\theta_3^4(i\tau_0)\theta_4^4(i\tau_0)]}, \nonumber\\
{\rm K}_{6}^{\frac{1}{2},0}(i\tau_0)&=&-\textstyle{\frac{1}{84}
[\theta_3^4(i\tau_0)+\theta_2^4(i\tau_0)][\frac{31}{16}\,\theta_4^8(i\tau_0)+
\theta_3^4(i\tau_0)\theta_2^4(i\tau_0)]}, \nonumber\\
~&\vdots&~\nonumber
\end{eqnarray}
Note that when $\tau_0\to\infty$ we have limits
$\theta_4(i\tau_0)\to 1$, $\theta_3(i\tau_0)\to 1$ and
$\theta_2(i\tau_0)\to 2 \exp{\left(-\frac{\pi\tau_0}{4}\right)}
\to 0$, and Kronecker's double series reduce to the Bernoulli
polynomials:
$$
\lim_{\tau_0\to\infty}{\rm
K}_{2p}^{\alpha,\beta}(i\tau_0)=B_{2p}^{\alpha}.
$$
The case $\tau_0 \to 0$ can be obtained by using Jacobi's
imaginary transformation of the $\theta(i\tau_0)$ - functions. In
this case $\theta_2(i\tau_0)\to \frac{1}{\sqrt{\tau_0}}$,
$\theta_3(i\tau_0)\to\frac{1}{\sqrt{\tau_0}}$ and
$\theta_4(i\tau_0)\to
\frac{2}{\sqrt{\tau_0}}\exp{\left(-\frac{\pi}{4\tau_0}\right)} \to
0$  and the Kronecker's function can again be reduce to the
Bernoulli polynomials.

\section{Jacobi transformation}
\label{jacobiTransformation}

We also need the behavior of the $\theta$ functions, Dedekind's
$\eta$-function and the Kronecker functions $K_{2p}^{0, 1/2}$
under the Jacobi transformation
\begin{eqnarray}
\tau \rightarrow \tau' = -1/\tau.
\label{def_Jacobi_transformation}
\end{eqnarray}
The result for the $\theta$ functions and Dedekind's
$\eta$-function when $z=0$ is given in ref.~\cite{Korn}
\begin{eqnarray}
\theta_{3}(\tau')   &=& (-i\tau)^{1/2} \theta_{3}(\tau), \nonumber\\
\theta_{2}(\tau') &=& (-i\tau)^{1/2} \theta_{4}(\tau), \label{def_Jacobi_transformation1}\\
\theta_{4}(\tau') &=& (-i\tau)^{1/2} \theta_{2}(\tau),\nonumber\\
\eta(\tau') &=& (-i\tau)^{1/2} \eta(\tau).\nonumber
\end{eqnarray}
The result for the Kronecker functions $K_{2p}^{0, 1/2}$ and
$K_{2p}^{1/2,0}$ can be obtain from the relation between
coefficients in Laurent expansion of the Weierstrass function and
Kronecker functions (see Appendix F in \cite{izmailian2002}) and
is given by
\begin{eqnarray}
K_{2p}^{0,1/2}(\tau')   &=& \tau^{2p} K_{2p}^{1/2,0}(\tau),
\label{def_Jacobi_transformation1Kron}\\
 K_{2p}^{1/2,0}(\tau') &=&
\tau^{2p} K_{2p}^{0,1/2}(\tau), \nonumber
\end{eqnarray}
In particular,  in the case of pure imaginary aspect ratio
$\tau=i\tau_0$, the $\theta$-functions and $K_{2p}^{0, 1/2}$,
$K_{2p}^{1/2,0}$ -functions transforms under
(\ref{def_Jacobi_transformation}) as follows
\begin{eqnarray}
\theta_{3}\left(i/\tau_0\right)
&=& \tau_0^{1/2} \theta_{3}(i\tau_0), \nonumber \\
\theta_{2}\left(i/\tau_0\right)
&=& \tau_0^{1/2} \theta_{4}(i\tau_0),\label{def preob}\\
\theta_{4}\left(i/\tau_0\right) &=&
\tau_0^{1/2} \theta_{2}(i\tau_0), \nonumber\\
\eta\left(i/\tau_0\right) &=& \tau_0^{1/2} \eta(i\tau_0),
\nonumber
\end{eqnarray}
\begin{eqnarray}
K_{2p}^{0,1/2}\left(i/\tau_0\right) &=& (i\tau_0)^{2p}
K_{2p}^{1/2,0}(i\tau_0),
\label{def_Jacobi_transformation2Kron}\\
K_{2p}^{1/2,0}\left(i/\tau_0\right) &=& (i\tau_0)^{2p}
K_{2p}^{0,1/2}(i\tau_0), \nonumber
\end{eqnarray}

\end{document}